\documentclass[10pt,conference]{IEEEtran}
\IEEEoverridecommandlockouts
\usepackage{cite}
\usepackage{amsmath,amssymb,amsfonts}
\usepackage{algorithmic}
\usepackage{graphicx}
\usepackage{textcomp}
\usepackage{xcolor}
\usepackage{amsmath}
\usepackage{amsfonts}
\usepackage{amssymb}
\usepackage{graphicx}
\usepackage{float}
\usepackage{gensymb}
\def\BibTeX{{\rm B\kern-.05em{\sc i\kern-.025em b}\kern-.08em
    T\kern-.1667em\lower.7ex\hbox{E}\kern-.125emX}}
\begin{document}

\title{On the Reliability of Estimation Bounds in Low-SNR Bistatic ISAC}

\author{\IEEEauthorblockN{Ataher Sams and Besma Smida}
\IEEEauthorblockA{Department of Electrical and Computer Engineering, University of Illinois  Chicago, USA\\
Emails: \{asams3, smida\}@uic.edu}
}

\maketitle

\begin{abstract}
    This paper explores a bistatic Integrated Sensing and Communication (ISAC) framework, where a base station transmits communication signal that serve both direct communication with a user and multi-target parameter estimation through reflections captured by a separate sensing receiver. We assume that the instantaneous knowledge of the transmit signal at the sensing receiver is not available, and the sensing receiver only has knowledge of the statistical properties of the received signal. Unlike prior research that focuses on power allocation or optimal beamforming design for ISAC, we emphasize the inadequacy of the Cramér-Rao Bound (and its variant) in low Signal-to-Noise Ratio (SNR) regimes, particularly in passive sensing scenarios. Due to severe path loss and other impairments, the received sensing SNR is often significantly lower than that of direct Line-of-Sight communication, making CRB-based performance evaluation unreliable. To address this, we adopt the Ziv-Zakai Bound (ZZB) for Angle of Arrival estimation, which provides a more meaningful lower bound on estimation error. We derive analytical expressions for the ZZB and the achievable ergodic communication rate as functions of SNR. Through numerical simulations, we analyze the pareto-front between communication and sensing performance, demonstrating why ZZB serves as a better metric in low sensing SNR ISAC where traditional CRB-based approaches fail.
\end{abstract}

\begin{IEEEkeywords}
bistatic, Integrated Sensing and Communication, CRB, ZZB
\end{IEEEkeywords}

\section{Introduction}
The increasing demand for high-speed wireless communication and precise sensing capabilities has driven the development of Integrated Sensing and Communication (ISAC) systems \cite{intro2}. Previous work on ISAC has predominantly focused on monostatic systems \cite{rel_2, rel_5}, where a single base station is responsible for both transmitting and receiving signals. The authors in \cite{rel_10} have examined the core compromises in ISAC operating over Gaussian channels, particularly focusing on how resources are allocated between subspace and deterministic-random approaches. In the study \cite{rel_11}, a communication-assisted sensing framework is introduced, utilizing rate-distortion theory to characterize the information-theoretic limits of ISAC systems and enhance sensing performance. In contrast, bistatic ISAC systems are more practical, as they do not require knowledge of the instantaneous transmit signal, necessitating the exploration of joint sensing and communication trade-offs in bistatic configurations. In this setup, the base station handles the transmission of signals, while user devices are responsible for receiving reflected signals and performing both communication and sensing tasks. Several studies have extended the exploration of bistatic ISAC systems. Recent investigations have expanded the theoretical understanding of bistatic ISAC systems. Notable contributions \cite{x1, rel_1} include beamforming strategies for passive object localization, the interplay between spatial and temporal beamforming, and the development of performance bounds for bistatic ISAC systems. Particularly relevant work \cite{x4} has focused on performance limitations in CSI-ratio-based bistatic Doppler sensing, providing closed-form derivations of bounds for Doppler estimation and analyzing the impact of physical parameters on system performance. 

Our research focuses on bistatic ISAC systems in multi-target scenarios, with particular emphasis on Angle of Arrival (AoA) estimation under low-SNR conditions. While most prior works on ISAC performance analysis rely on the Cramér–Rao Bound (CRB) or its variants—such as the Bayesian CRB (BCRB) and the Ergodic CRB (ECRB)—these approaches are primarily effective in high-SNR regimes, where the CRB serves as a reliable benchmark. However, in low-SNR scenarios, the CRB becomes overly optimistic and fails to provide meaningful insights, since it is an unbiased estimator and does not properly incorporate prior information on the parameters of interest. This limitation motivates the adoption of a more robust performance metric. In our work, we employ the Ziv–Zakai Bound (ZZB), which offers several advantages over the CRB. Unlike the CRB, the ZZB requires only mild regularity conditions and can naturally integrate prior probability density functions of the estimation parameters. More importantly, the ZZB is known for its tightness in high-noise regimes, making it particularly suitable for characterizing performance in low-SNR ISAC systems. Despite its seemingly sophisticated formulation, ZZB can be applied to a wide range of estimation tasks, such as position and AoA estimation \cite{zzb_mimo_1, zzb4}. Recent research has further highlighted the usefulness of the ZZB in ISAC applications. For instance, Sun et al. \cite{zzb_IPAC} investigated the positioning–communication trade-off by applying the ZZB to optimize power allocation in an IPAC system, while Miao et al. \cite{zzb_near_field} derived both CRB and ZZB expressions for joint position and velocity estimation in near-field ISAC scenarios under Doppler and spatial wideband effects. In parallel, Zhang et al. \cite{zzb3} advanced the analysis of ZZB by deriving an explicit form for multi-target AoA estimation across a wide SNR range. Our work extends Zhang’s formulation to the bistatic ISAC setting, where both communication and sensing coexist. By capturing this fundamental trade-off using ZZB, our analysis provides a more realistic and reliable benchmark for evaluating multi-target AoA estimation in bistatic ISAC systems operating under low-SNR conditions.

Although the ZZB is well studied in array processing, it remains largely unexplored in ISAC. The main contributions of this work can be summarized as follows: We analyze the Ziv-Zakai Bound for bistatic ISAC systems, establishing a theoretical framework that characterizes the fundamental limits of AoA performance in such systems. Through performance bound analysis, we demonstrate the limitations of conventional Cramér-Rao Bound in bistatic ISAC systems, particularly in low SNR scenarios. Our results reveal that ZZB provides significantly more realistic estimation bounds than CRB by accounting for threshold effects and prior information about target locations. We present a rate-sensing trade-off study based on subspace joint beamforming that challenges the reliability of CRB-based analysis in low passive sensing SNR regimes, showing that traditional CRB-based trade-off curves can be overly optimistic and potentially misleading.

\section{System Model} \label{section2}

\begin{figure} [H]
    \centering
    \includegraphics[width=7cm, height=4cm]{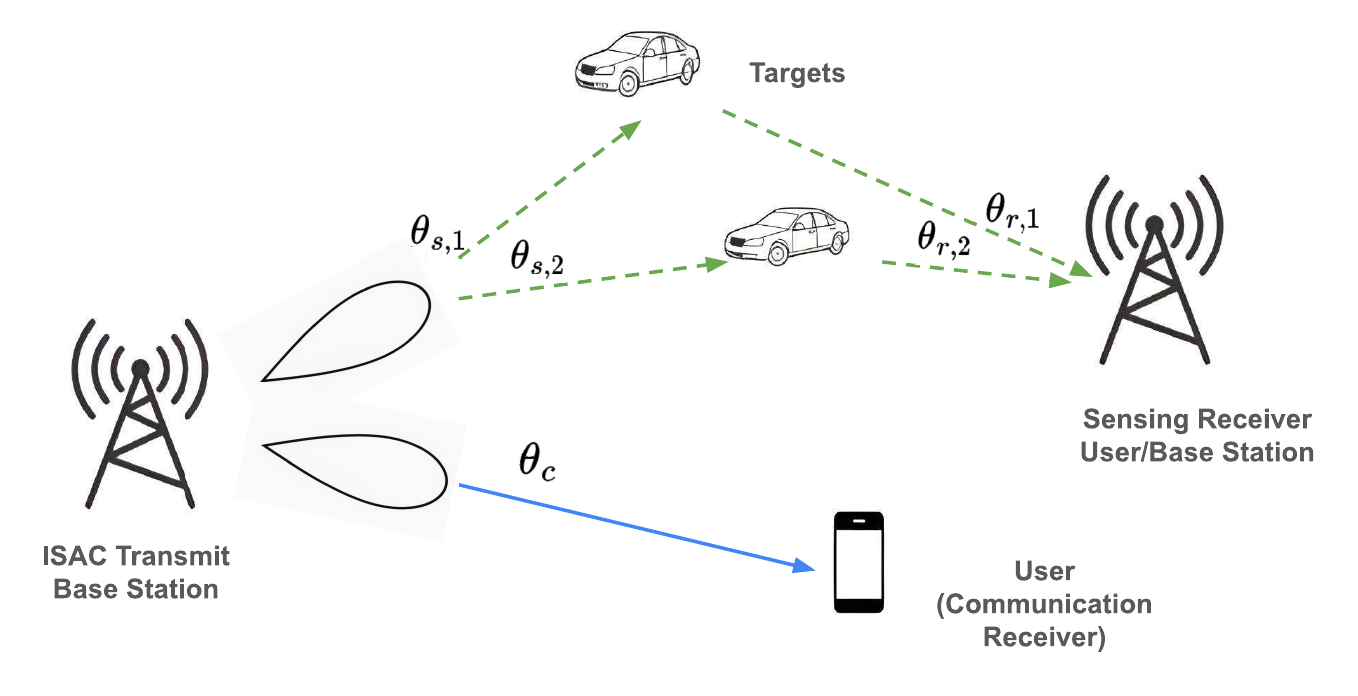}
    \caption{Bistatic Integrated Sensing and Communication (ISAC) system with MIMO setup. The transmitter maintains communication with a Communication RX (CRx), while the Sensing RX (SRx) detects targets through reflected signals. }
    \label{fig:enter-label}
\end{figure}

We consider a bistatic Integrated Sensing and Communication (ISAC) system, where the transmitter (TX) simultaneously serves communication and radar sensing functionalities. Specifically, the TX and the passive sensing receiver (SRx) are each equipped with Uniform Linear Arrays (ULAs) consisting of $M_{\mathrm{Tx}}$ and $M_{\mathrm{Rx}}$ antenna elements, respectively. A separate communication receiver (CRx) is positioned at a known angle $\theta_c$ relative to the transmitter. The TX transmits a single-stream communication signal over $L$ observation snapshots. The transmitted waveform vector for the single-stream scenario at snapshot $\ell$ is denoted by $s^{(\ell)} \in \mathbb{C}$, with $\ell=1,2, \ldots, L$. These waveforms are assumed independent, identically distributed (i.i.d.), zero-mean circularly symmetric complex Gaussian random variables with unit variance, i.e., $\mathbb{E}\left\{\left|s^{(\ell)}\right|^2\right\}=1$. The collection of these transmitted symbols across $L$ snapshots is represented by the waveform vector $\mathbf{s}=$ $\left[s^{(1)}, s^{(2)}, \ldots, s^{(L)}\right]^T \in \mathbb{C}^{L \times 1}$, having covariance structure $\mathbf{R}_s=\mathbb{E}\left\{\mathbf{s s}^H\right\}=\mathbf{I}_L$. The single transmitted data stream is shaped spatially by a fixed transmit beamforming vector $\mathbf{w} \in$ $\mathbb{C}^{M_{\mathrm{Tx}} \times 1}$, resulting in a transmit signal vector for each snapshot given by:
\begin{equation}
\mathbf{x}^{(\ell)}=\mathbf{w} s^{(\ell)} \in \mathbb{C}^{M_{\mathrm{Tx}} \times 1}
\end{equation}

\subsection{Sensing Model}
We now focus on the sensing subsystem at the sensing receiver (SRx), which passively observes reflections from $K$ point-like targets situated in the far-field. Each target $k, k=1,2, \ldots, K$, is characterized by a reflection coefficient $\gamma_k \in \mathbb{C}$, representing both the radar cross-section (RCS) and propagation-induced attenuation, an angle of departure (AoD) $\theta_{s, k}$ from the TX, and an angle of arrival (AoA) $\theta_{r, k}$ at the SRx .
For the $\ell$-th snapshot, the received signal vector $\mathbf{y}^{(\ell)} \in \mathbb{C}^{M_{\mathrm{Rx}} \times 1}$ at the $\operatorname{SRx}$ is modeled explicitly as:
\begin{equation}
\mathbf{y}^{(\ell)}=\sum_{k=1}^K \gamma_k \mathbf{a}_{\mathrm{Rx}}\left(\theta_{r, k}\right) \mathbf{a}_{\mathrm{Tx}}^H\left(\theta_{s, k}\right) \mathbf{w} s^{(\ell)}+\boldsymbol{z}^{(\ell)}
\end{equation}
where $\mathbf{a}_{\mathrm{Rx}}\left(\theta_{r, k}\right) \in \mathbb{C}^{M_{\mathrm{Rx}} \times 1}$ and $\mathbf{a}_{\mathrm{Tx}}\left(\theta_{s, k}\right) \in \mathbb{C}^{M_{\mathrm{Tx}} \times 1}$ denote the normalized receive and transmit steering vectors for the SRx and TX antenna arrays, respectively. Both steering vectors correspond to ULAs with half-wavelength antenna spacing. The additive noise at snapshot $\ell$, denoted by $\boldsymbol{z}^{(\ell)} \in$ $\mathbb{C}^{M_{\mathrm{Rx}} \times 1}$, is assumed independent across snapshots and modeled as complex Gaussian with zero mean and covariance $\sigma^2 \mathbf{I}_{M_{\mathrm{Rx}}}$, i.e., $\boldsymbol{z}^{(\ell)} \sim \mathcal{C N}\left(0, \sigma^2 \mathbf{I}_{M_{\mathrm{Rx}}}\right)$

Aggregating observations over all $L$ snapshots, we form the received signal matrix, $\mathbf{Y}=\left[\mathbf{y}^{(1)}, \mathbf{y}^{(2)}, \ldots, \mathbf{y}^{(L)}\right] \in \mathbb{C}^{M_{\mathrm{Rx}} \times L}$ which can be written compactly as:
\begin{equation}
\mathbf{Y}=\sum_{k=1}^K \gamma_k \mathbf{a}_{\mathrm{Rx}}\left(\theta_{r, k}\right) \mathbf{a}_{\mathrm{Tx}}^H\left(\theta_{s, k}\right) \mathbf{w s}^T+\mathbf{Z}
\end{equation}
where $\mathbf{s}=\left[s^{(1)}, s^{(2)}, \ldots, s^{(L)}\right]^T \in \mathbb{C}^{L \times 1}$ and $\mathbf{Z}=\left[\boldsymbol{\zeta}^{(1)}, \ldots, \boldsymbol{\zeta}^{(L)}\right] \in \mathbb{C}^{M_{\mathrm{Rx}} \times L}$.
In contrast to traditional active radar or ISAC scenarios, where the transmitted signal waveforms are explicitly known to the sensing receiver- our scenario adopts a passive target estimation paradigm. Hence, we assume that the instantaneous transmitted signal waveform are unknown at the SRx. Instead, the receiver exploits statistical properties of the transmitted waveforms by estimating the covariance of the received signal. The transmitted waveform covariance matrix for this single-stream scenario is consequently defined as,
\begin{equation}
\mathbf{R}_x=\mathbb{E}\left\{\mathbf{x}(t) \mathbf{x}(t)^H\right\}=\mathbf{w} \mathbf{w}^H
\end{equation}
Therefore, the covariance matrix of the received signals at SRx denoted as, $\mathbf{R}_y \in \mathbb{C}^{M_{\mathrm{Rx}} \times M_{\mathrm{Rx}}}$, takes the following explicit stochastic form,
\begin{equation}
\begin{split}
\mathbf{R}_y = \sum_{k=1}^K |\gamma_k|^2 \left( \mathbf{a}_{\mathrm{Tx}}^H(\theta_{s, k}) \mathbf{w} \mathbf{w}^H \mathbf{a}_{\mathrm{Tx}}(\theta_{s, k}) \right) \\\mathbf{a}_{\mathrm{Rx}}(\theta_{r, k}) \mathbf{a}_{\mathrm{Rx}}^H(\theta_{r, k}) 
+ \sigma^2 \mathbf{I}_{M_{\mathrm{Rx}}}
\end{split}
\end{equation}

\subsection{Communication System Model}

The downlink communication channel from the ISAC transmitter to the single-antenna communication receiver (CRx) is modeled as
\begin{equation}
\mathbf{h}_{\text{DL}} = \alpha_c \mathbf{a}_{\text{Tx}}^H(\theta_c) \in \mathbb{C}^{1 \times M_{\text{Tx}}},
\end{equation}
where $\alpha_c \in \mathbb{C}$ denotes the LOS channel coefficient, and $\mathbf{a}_{\text{Tx}}(\theta_c)$ is the transmit steering vector towards the CRx.\\
At snapshot $\ell$, the received signal at the CRx is
\begin{equation}
y_c^{(\ell)} = \alpha_c \mathbf{a}_{\text{Tx}}^H(\theta_c)\mathbf{w}s^{(\ell)} + z_c^{(\ell)},
\end{equation}
where the additive noise is $z_c^{(\ell)}\sim\mathcal{CN}(0,\sigma_c^2)$, the transmit beamforming vector $\mathbf{w}$ satisfies the power constraint $\|\mathbf{w}\|^2 = P_t$.
The downlink ergodic achievable rate is given by,
\begin{equation}
R_{\text{in}} = \mathbb{E}_{\alpha_c}\left\{\log_2\left(1+|\alpha_c|^2|\mathbf{a}_{\text{Tx}}^H(\theta_c)\mathbf{w}|^2/{\sigma_c^2}\right)\right\},
\end{equation}
where the expectation is taken over the channel coefficient $\alpha_c$, which is assumed known at the receiver.

\section{Estimation Bounds for Estimating AoA} \label{section3}
The Cramér-Rao Bound (CRB) effectively establishes a minimum variance for unbiased estimators in high signal-to-noise ratio (SNR) situations where estimation errors remain small. However, this bound becomes less useful and unrealistically optimistic when dealing with low SNR conditions. This limitation stems from two key assumptions in the CRB's derivation: the requirement for unbiased estimation and the absence of prior parameter information. The Ziv-Zakai Bound (ZZB) takes a different approach by incorporating the parameter's prior distribution into its calculations. This inclusion of prior information allows the ZZB to maintain its validity and provide more accurate performance predictions across all SNR levels, including challenging low SNR environments where the CRB's assumptions no longer hold. Consequently, the ZZB proves more practical for real-world applications where noise might be more dominant.

\subsection{Ziv-Zakai Bound}
For our bistatic ISAC system with multiple targets, we derive the ZZB to establish fundamental limits on the AoA estimation accuracy. Let the parameter vector $\boldsymbol{\Theta}_r = [\theta_{r,1}, \theta_{r,2}, \ldots, \theta_{r,K}]^T$ represent the AoAs of $K$ targets at the sensing receiver (SRx), where $\theta_{r,k}$ corresponds to the same targets discussed in our system model. We assume $\boldsymbol{\Theta}_r$ is a random vector with a known prior probability density function $p(\boldsymbol{\Theta}_r)$. The ZZB establishes a fundamental relationship between estimation accuracy and detection performance by relating the estimation problem to a series of binary hypothesis detection tests:
\begin{align}
   H_0 &: \text{Target AoAs at } \boldsymbol{\Theta}_r, & \mathbf{Y} &\sim p(\mathbf{Y} \mid \boldsymbol{\Theta}_r), \\
   H_1 &: \text{Target AoAs at } \boldsymbol{\Theta}_r + \boldsymbol{\delta}, & \mathbf{Y} &\sim p(\mathbf{Y} \mid \boldsymbol{\Theta}_r + \boldsymbol{\delta}).
\end{align}
For any estimator $\hat{\boldsymbol{\Theta}}_r(\mathbf{Y})$ based on our previously defined received signal matrix $\mathbf{Y}$, the estimation error vector is defined as $\boldsymbol{\epsilon} = \hat{\boldsymbol{\Theta}}_r(\mathbf{Y}) - \boldsymbol{\Theta}_r$
with corresponding error covariance matrix, $\mathbf{R}_{\boldsymbol{\epsilon}} = \mathbb{E}\left\{ \boldsymbol{\epsilon} \boldsymbol{\epsilon}^T \right\}$
where the expectation is taken with respect to the joint pdf $p(\mathbf{Y}, \boldsymbol{\Theta}_r) = p(\mathbf{Y} \mid \boldsymbol{\Theta}_r) p(\boldsymbol{\Theta}_r)$. For any vector $\mathbf{a}$, the bound on the weighted mean square error is given by,
\begin{equation} \label{zzb_original}
 \text{ZZB}(\boldsymbol{\Theta}_{\mathbf{r}}) = \mathbf{a}^T \mathbf{R}_{\boldsymbol{\epsilon}} \mathbf{a} \geq \int_{0}^{\infty} \mathcal{V} \left\{ \max_{\boldsymbol{\delta}: \mathbf{a}^T \boldsymbol{\delta} = h} A(\boldsymbol{\delta}) P_{\text{min}}(\boldsymbol{\delta}) \right\} h \, \mathrm{d}h,
\end{equation}
where $\mathcal{V}\{\cdot\}$ represents the valley-filling operation, and
$
   A(\boldsymbol{\delta}) = \int \min \left[ p(\boldsymbol{\Theta}_r), p(\boldsymbol{\Theta}_r + \boldsymbol{\delta}) \right] \mathrm{d}\boldsymbol{\Theta}_r,
$
quantifies the overlap between the a priori parameter distributions, while $p(\boldsymbol{\Theta}_r)$ and $p(\boldsymbol{\Theta}_r + \boldsymbol{\delta})$ are a priori pdf of the AoA. As demonstrated in previous research \cite{zzb1,zzb3}, the valley-filling operation does not significantly affect the AoA estimation context. The expression at \eqref{zzb_original} can be reformulated as,
\begin{align}
\text{ZZB}\left(\boldsymbol{\Theta}_{\mathbf{r}}\right) 
\geq \frac{1}{2} \int_0^{\infty} 
\max_{\boldsymbol{\delta}: \mathbf{a}^T \boldsymbol{\delta} = h} &
\left[ \int_{\mathbb{R}^K} 
\left( p\left(\boldsymbol{\Theta}_r\right) + p\left(\boldsymbol{\Theta}_r + \boldsymbol{\delta}\right) \right) \right. \notag \\
&\quad \left. \times P_{\min}(\boldsymbol{\delta}) \, d\boldsymbol{\Theta}_r \right] h \, dh
\label{eq:multi_target_zzb}
\end{align}
Under our setting, we can consider our hypothesis problem as a Binary symmetric problem, for which we can approximate $P_{min}(\boldsymbol{\delta})$ \cite[p71]{zzb2},
\begin{equation}\label{equ_p}
\begin{aligned}
P_{\min}(\boldsymbol{\delta})
&\approx
Q\!\left(
\sqrt{\frac{\partial^2 \mu\!\left(p;\boldsymbol{\delta}\right)}{\partial p^2}}
\right)
\exp\!\left(
\mu\!\left(p;\boldsymbol{\delta}\right)
+ \frac{1}{8}\frac{\partial^2 \mu\!\left(p;\boldsymbol{\delta}\right)}{\partial p^2}
\right).
\end{aligned}
\end{equation}
where $\mu(p; \boldsymbol{\delta})$ is the semi-invariant moment generating function, defined as $\mu(p; \boldsymbol{\delta}) = \ln \int f(\mathbf{Y} \mid \boldsymbol{\Theta}_r + \boldsymbol{\delta})^p f(\mathbf{Y} \mid \boldsymbol{\Theta}_r)^{1-p} \, \mathrm{d}\mathbf{Y}.
$ Since the two hypotheses have equal prior probabilities, we set $p=\frac{1}{2}$, and evaluate the semi-invariant moment generating function $\mu(p ; \boldsymbol{\delta})$ at this value. For this evaluation, we define $\boldsymbol{R}_{\mathbf{Y} \mid \boldsymbol{\Theta}_r}$ as the covariance matrix of $\mathbf{Y}$ conditioned on $\boldsymbol{\Theta}_r$, and $\boldsymbol{R}_{\mathbf{Y} \Theta_r+\boldsymbol{\delta}}$ as the covariance matrix conditioned on $\boldsymbol{\Theta}_r+\boldsymbol{\delta}$. For simplicity, we use $\boldsymbol{R}$ and $\boldsymbol{R}_{\boldsymbol{\delta}}$ to represent these matrices, respectively. The expressions for $\mu(p; \boldsymbol{\delta})$ and its second derivative at $p = \frac{1}{2}$ are then given by \cite{zzb3},

\begin{align}
& \left.\mu\left(p ; \boldsymbol{\delta}\right)\right|_{p=\frac{1}{2}} =
T\left[\frac{\ln\left(|\boldsymbol{R}| |\boldsymbol{R}_{\boldsymbol{\delta}}|\right)}{2} - \ln\left|\frac{\boldsymbol{R} + \boldsymbol{R}_{\boldsymbol{\delta}}}{2}\right|\right]
\label{equ_p_mu} \\[0.5em]
& \left.\frac{\partial^2 \mu\left(p ; \boldsymbol{\delta}\right)}{\partial p^2}\right|_{p=\frac{1}{2}} =
4T \, \operatorname{Tr}\left\{ \left( (\boldsymbol{R} + \boldsymbol{R}_{\boldsymbol{\delta}})^{-1} (\boldsymbol{R} - \boldsymbol{R}_{\boldsymbol{\delta}}) \right)^2 \right\}
\label{equ_p_dmu}
\end{align}
$P_{min}(\boldsymbol{\delta)}$ in \eqref{P_min_delta} behaves differently in high and low sensing SNR scenario based on pertubation $\boldsymbol{\delta}$,
\begin{align} \notag
P_{min}(\boldsymbol{\delta}) \approx 
\begin{cases}
P_{\min,m}(\boldsymbol{\delta}), & \text{(High SNR, error depends on } \boldsymbol{\delta} \text{)} \\
P_{\min,n}, & \text{(Low SNR, constant error floor)}
\end{cases}
\label{eq:P_delta_case}
\end{align}
For small perturbations $\boldsymbol{\delta}$ (corresponding to the low estimation error region at high sensing SNR), we can perform a second-order Taylor series expansion of equations \eqref{equ_p_mu} and \eqref{equ_p_dmu} around $\boldsymbol{\delta} \approx \mathbf{0}$. This yields an approximation of $P_{\min}$ for the mainlobe region (around $\delta$):
\begin{equation} \label{P_min_delta}
P_{\min,m}(\boldsymbol{\delta}) \approx Q\left(\frac{1}{2} \sqrt{\boldsymbol{\delta}^T \mathbf{CRB}(\boldsymbol{\Theta}_r)^{-1}\boldsymbol{\delta}}\right)
\end{equation}
However, for larger perturbations $\boldsymbol{\delta}$ (corresponding to the high estimation error regime at low sensing SNR), the approximation in equation \eqref{P_min_delta} breaks down. In this case, we need to account for the influence of array ambiguities that occur at the nulls of the beampattern. Following the derivation in \cite[Appendix A]{zzb3}, the minimum probability of error $P_{\min,n}$ at these nulls can be expressed as \cite{zzb3}:
\begin{equation} \label{P_min,n,multi}
\small
\begin{split} 
&P_{min,n} =  \mathcal{Q} \left( \sqrt{2 L \sum_{k=1}^K \left( \frac{M_{\mathrm{Rx}} \eta_{s, k}}{2+M_{\mathrm{Rx}} \eta_{s, k}} \right)^2 } \right) \cdot  \\
&\exp \left( L \sum_{k=1}^K \left[ \ln \left( \frac{4\left(1+M_{\mathrm{Rx}} \eta_{s, k}\right)}{\left(2+M_{\mathrm{Rx}} \eta_{s, k}\right)^2} \right) 
+ \left( \frac{M_{\mathrm{Rx}} \eta_{s, k}}{2+M_{\mathrm{Rx}} \eta_{s, k}} \right)^2 \right] \right)
\end{split}
\end{equation}
where $\eta_{s,k}$ represents the sensing SNR for the $k$-th target in the bistatic configuration, defined as:
$\eta_{s,k} = |\gamma_k|^2 |\mathbf{a}_{\mathrm{Tx}}^H(\theta_{s,k}) \mathbf{w}|^2 / {\sigma^{2}}$.

Under the assumption of orthogonal transmitted signals and independent target echoes (which can be maintained by ensuring targets are separated beyond a single equivalent beamwidth), we can simplify the general formula in equation \eqref{zzb_original} which would otherwise require grid search and numerical integration. Following the approach in \cite{zzb4, zzb3}, we develop a more tractable expression based on known a priori variance and CRB. By applying Sylvester’s determinant theorem and the Woodbury matrix identity, as shown in \cite{zzb3}, the ZZB formulation results in a tighter bound compared to earlier approaches in \cite{zzb1, zzb_mimo_1},
\begin{equation} \label{zzb_final}
\small
\mathrm{ZZB}(\boldsymbol{\Theta}_r) \geq 
\underbrace{2 P_{\min,n} \mathcal{B}_{\mathrm{AP}}}_{\begin{array}{c}
\text{\scriptsize for high estimation} \\
\text{\scriptsize error region}
\end{array}} + 
\underbrace{\Gamma_{3/2}\left(\tilde{u}\right) 
\frac{\operatorname{Tr}\{\mathrm{CRB}(\boldsymbol{\Theta}_r)\}}{K}}_{\begin{array}{c}
\text{\scriptsize for low estimation} \\
\text{\scriptsize error region}
\end{array}}
\end{equation}
where $\mathrm{CRB}(\boldsymbol{\Theta}_r)$ is the Cramér-Rao Bound matrix as defined later in equation \eqref{crb}, which properly accounts for the bistatic configuration.
where, The first term $2 P_{\min,n} \mathcal{B}_{\mathrm{AP}}$ represents the contribution from the a priori region, with $\mathcal{B}_{\mathrm{AP}}$ being the a priori bound derived from the order statistics of the uniformly distributed AoAs over range $\zeta$.
The second term emerges from the analytical evaluation of the integral in equation \eqref{zzb_original} for small estimation errors. The incomplete gamma function $\Gamma_{3/2}(\tilde{u})$ stems by solving the optimization problem in the integrand for the binary hypothesis detection problem from \cite[Appendix B]{zzb3}, 
\begin{equation}
\max_{\boldsymbol{\delta}: \mathbf{a}^T \boldsymbol{\delta} = h} P_{\min}(\boldsymbol{\delta}) \approx Q\left(\frac{\sqrt{K}h}{2\sqrt{\mathbf{1}_K^T\mathrm{CRB}(\boldsymbol{\Theta}_r)\mathbf{1}_K/K}}\right)
\end{equation}
The parameter $\tilde{u}$ serves as the upper limit of integration and determines the transition between the a priori (higher estimation error region) and asymptotic regions (lower estimation error region) for our bistatic ISAC system.

This ZZB formulation in \eqref{zzb_final} provides several key insights into AoA estimation performance for our bistatic ISAC system. In high SNR conditions (asymptotic region), $\Gamma_{3/2}(\tilde{u}) \approx 1$ and the bound approaches the CRB, $\operatorname{Tr}\{\mathrm{CRB}(\boldsymbol{\Theta}_r)\}/{K}$. In low SNR conditions (a priori region), $\Gamma_{3/2}(\tilde{u}) \approx 0$ and $2P_{\min,n} \approx 1$, causing the bound to converge to the a priori bound $\mathcal{B}_{\mathrm{AP}}$. In the threshold region between these extremes, both terms contribute to the bound.

\subsection{Cramér-Rao Bound}
In order to evaluate \eqref{zzb_final}, we will need to find out the CRB of the AoA of our setting. Defining the receive steering matrix $\mathbf{A}_{\mathrm{Rx}}=\left[\mathbf{a}_{\mathrm{Rx}, 1}, \mathbf{a}_{\mathrm{Rx}, 2}, \ldots, \mathbf{a}_{\mathrm{Rx}, K}\right] \in \mathbb{C}^{M_{\mathrm{Rx}} \times K}$, and $\boldsymbol{\Gamma}_x$ a diagonal matrix collecting the effective signal power from each target; covariance $\mathbf{R}_y$ can be expressed in a compact form-  $\mathbf{R}_y=\mathbf{A}_{\mathrm{Rx}} \boldsymbol{\Gamma}_x \mathbf{A}_{\mathrm{Rx}}^H+\sigma^2 \mathbf{I}_{M_{\mathrm{Rx}}}$. As per our passive bistatic sensing scenario where the SRx has no direct knowledge of the instantaneous transmitted signal waveforms, the SRx has access only to the statistical properties of the received signals through the covariance matrix $\mathbf{R}_y$. As described before, the SRx observes signal reflections whose information content is treated as a random Gaussian process rather than a known deterministic quantity. From \cite{stochastic_CRB, x1}, we can utilized the CRB formulation for stochastic multi-target model, 
\begin{equation} \label{crb}
\operatorname{CRB}\left(\boldsymbol{\Theta}_r\right)=\frac{\sigma^2}{2 L}\left\{\Re\left[\mathbf{\Pi_H} \circ\left(\boldsymbol{\Gamma}_x \mathbf{A}_{\mathbf{R x}}^H \mathbf{R}_y^{-1} \mathbf{A}_{\mathbf{R x}} \boldsymbol{\Gamma}_x\right)^T\right]\right\}^{-1}
\end{equation}
where, $\mathbf{\Pi_H}=\dot{\mathbf{A}}_{\mathbf{R} \mathbf{x}}^H\left[\mathbf{I}-\mathbf{A}_{\mathbf{R} \mathbf{x}}\left(\mathbf{A}_{\mathbf{R} \mathbf{x}}^H \mathbf{A}_{\mathbf{R x}}\right)^{-1} \mathbf{A}_{\mathbf{R} \mathbf{x}}^H\right] \dot{\mathbf{A}}_{\mathbf{R x}}$ and $\dot{\mathbf{A}}_{\mathbf{R x}}$ is the derivative of $\mathbf{A}_{\mathbf{R x}}$ with respect to the parameters $\boldsymbol{\Theta}_r$. A note here is that we do not compare ZZB with the Bayesian Cramér-Rao Bound (BCRB) in Section \ref{section4} because parameters with uniform prior distributions do not satisfy the regularity conditions required for the BCRB \cite{bcrb}.

\section{Numerical Results} \label{section4}

\begin{figure}
    \centering
    \includegraphics[width=0.95\linewidth]{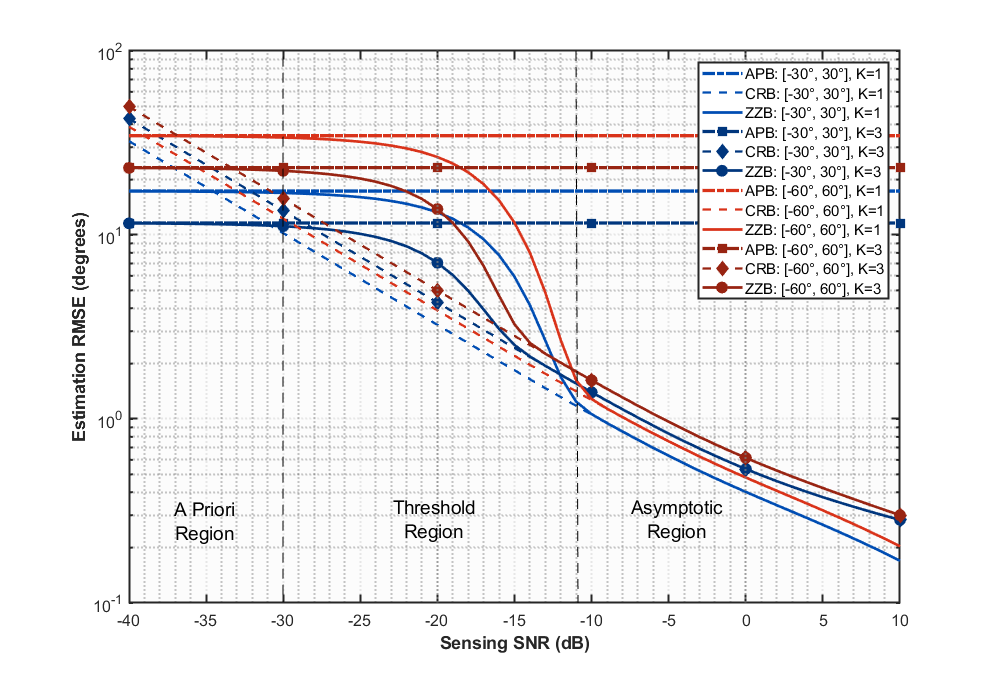}
    \caption{Performance comparison of Angle of Arrival (AoA) estimation bounds for different AoA ranges and targets. The plot shows the Root Mean Square Error (RMSE) in degrees versus Sensing SNR for Ziv-Zakai Bound (ZZB), Cramér-Rao Bound (CRB) and A Priori Bound (APB). Results are presented for two AoA ranges: [-30°, 30°] and [-60°, 60°] for single and three target cases.} 
    \label{fig:AoAestimation}
\end{figure}

In this section, we present the simulation results for using the Ziv-Zakai Bound as a reliable sensing metric. We consider a multi-target scenario where each AoA follows a uniform distribution. The simulation is conducted using $\ell = 500$ monte carlo trials for each receiver Signal-to-Noise Ratio (SNR) to obtain the Root Mean Square Error (RMSE) of the CRB, ZZB, and a Priori Bound (APB). We assume the AoA follows a uniform distribution, so the apriori variance will be equal to $\zeta^2/12$, where $\zeta$ range of AoA. And the APB $\mathcal{B}_{\mathrm{AP}}$ becomes $\frac{K \zeta^2}{(K+1)^2(K+2)}$. 
Consequently, we employ the stochastic CRB at equation \eqref{crb} for comparison in this analysis. The RMSE for multi-target estimation scenario is calculated using the formula, $\operatorname{RMSE}=\sqrt{\frac{1}{\mathcal{L} K} \sum_{\ell=1}^{\mathcal{L}} \sum_{k=1}^K\left(\hat{\theta}_{\ell,(k)}-\theta_{\ell,(k)}\right)^2}$
We consider two AoA ranges: [-30°, 30°] and [-60°, 60°]. The system parameters include the number of snapshots, L = 100, and both the ISAC Tx and sensing receiver (SRx) to employ a 8-element ULA with half-wavelength inter-element spacing. We consider for the simulation, $K$ targets all have same SNRs. In Fig. \ref{fig:AoAestimation}, the ZZB demonstrates superior performance as a practical bound. In line with the theory, ZZB converges with the CRB at higher SNR (above -10 dB) and approaches the A Priori Bound (APB) at very low SNR (below -30 dB). Crucially, the ZZB provides a tighter and more realistic bound than the CRB in the noise dominated region successfully capturing the threshold effect. The Maximum Likelihood Estimator will closely follow the ZZB, at all SNR levels, rather than CRB as shown in \cite{zzb1}. We consider optimal sensing beamforming for this simulation. This demonstrates the practical achievability of the ZZB and validates its use as a performance metric.

Now, we discuss how trade-off between communication and sensing performance works in terms of transmit beamforming. Unlike traditional studies addressing power allocation and optimal beamforming \cite{rel_1,rel_2}, our focus is to highlight the shortcomings of widely used CRB (or variant like BCRB or ECRB) in the low-SNR domain, particularly in passive sensing, where the received sensing SNR is inherently lower than direct LOS communication SNR due to path loss and other adverse effects. For clarity, we exclude the beamforming gain from the definitions of the Sensing and Communication SNRs for the subsequent simulations. This trade-off is governed by the design of the single beamforming vector \(\mathbf{w}_{\text{tradeoff}}\), which is formed as a weighted combination of the communication-optimal beamformer \(\mathbf{w}_{\text{comm}}\) and the sensing-optimal beamformer \(\mathbf{w}_{\text{sensing}}\). This approach, also referred to as Subspace Joint Beamforming (SJB), for a single sensing target and single communication user, can be mathematically represented as follows:
\begin{equation}
\mathbf{w}_{\text{tradeoff}} = \frac{ \alpha \cdot \mathbf{w}_{\text{comm}} + (1 - \alpha) \cdot \mathbf{w}_{\text{sensing}} }{ \left\| \alpha \cdot \mathbf{w}_{\text{comm}} + (1 - \alpha) \cdot \mathbf{w}_{\text{sensing}} \right\| }
\end{equation}
where \(\alpha \in [0,1]\) is a parameter that balances the trade-off between communication and sensing objectives. The communication-optimal beamformer, \(\mathbf{w}_{\text{comm}} = \frac{\mathbf{a}_{\text{Tx}}(\theta_c)}{\left\| \mathbf{a}_{\text{Tx}}(\theta_c) \right\|}\), is aligned with the Angle of Arrival (AoA) of the communication user (\(\theta_c\)), while the sensing-optimal beamformer, \(\mathbf{w}_{\text{sensing}} = \frac{\mathbf{a}_{\text{Tx}}(\theta_s)}{\left\| \mathbf{a}_{\text{Tx}}(\theta_s) \right\|}\), targets the AoA of the sensing target (\(\theta_s\)). By adjusting \(\alpha\), one can dynamically balance the two goals: setting \(\alpha = 1\) maximizes the communication rate but increases the AoA estimation error, while setting \(\alpha = 0\) minimizes the error at the expense of communication performance. Intermediate values of \(\alpha\) provide a trade-off curve that represents the Pareto front of both objectives. A note here is that, any alternative beamforming strategies will also yield similar results.

\begin{figure}[t]
    \centering    \includegraphics[width=0.95\linewidth]{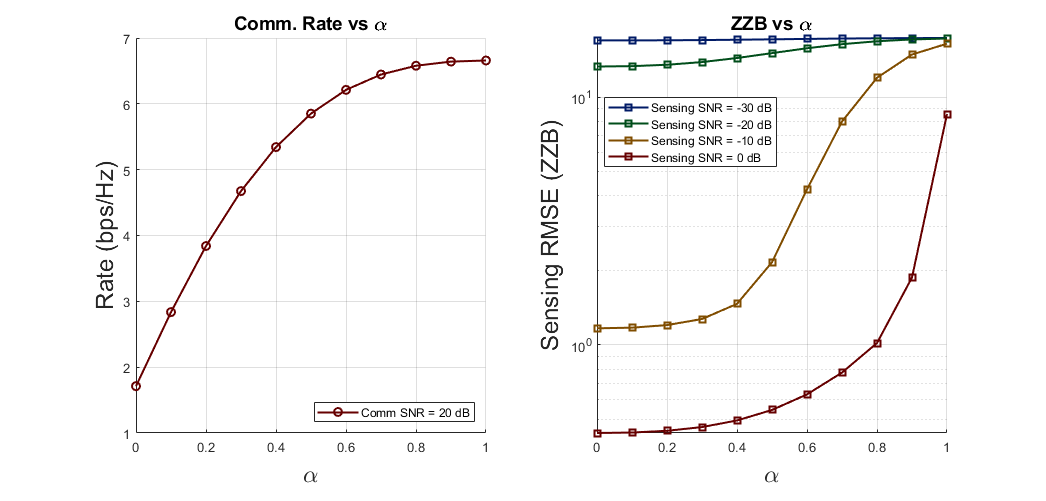}
    \caption{\textit{Left:} Communication rate (bps/Hz) versus beamforming parameter $\alpha$, showing how steering more power toward the user ( $\alpha \rightarrow 1$ ) increases rate. Communication SNR is fixed at 20 dB. 
\textit{Right:} Sensing estimation error (ZZB, in degrees) versus $\alpha$, for sensing SNRs of -30 , -20, -10 and 0 dB.}
    \label{fig:fig3}
\end{figure}
\begin{figure} 
    \centering    \includegraphics[width=0.95\linewidth]{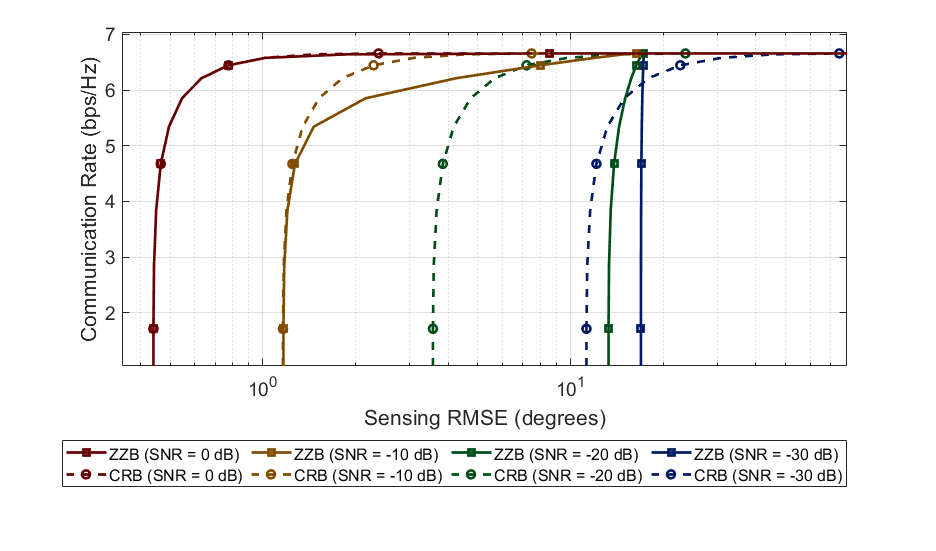}
    \caption{Comparison of communication rate and sensing performance bounds (CRB and ZZB). Notably, in the low sensing SNR regime (-20 dB and -30 dB), The ZZB curves reveal that beamforming strategies have limited impact on estimation performance.}
    \label{fig:fig4}
\end{figure}

Now assuming the communication SNR is fixed at 20 dB, varying the sensing SNR from low to high reveals the trade-off behavior in Fig. \ref{fig:fig3} and Fig. \ref{fig:fig4}. We arbitrarily set $\theta_s = 5\degree$, $\theta_r= 15\degree$ and $\theta_c =45\degree$. In Fig. \ref{fig:fig3}, we demonstrate the communication rate against $\alpha$ in sensing optimal and communication optimal case, where we can see gradual increase of communication rate as $\alpha$ forces the beamforming to be aligned to the communication user (CRx). While for ZZB vs $\alpha$ plot, it can be seen, the change in the bound depends highly on transmit beamforming configuration at higher SNR. In Fig. \ref{fig:fig4}, we provide a comparison of Communication rate vs the bounds. At a higher sensing SNR of 0 dB or above, both the CRB and ZZB converge, yielding an identical trade-off curve. This convergence indicates that at higher SNR levels, the ZZB provides the same estimation capability as the CRB. When the sensing SNR is around the boundary of the threshold region (-10 dB), the trade-off behavior exhibits a notable sensitivity to changes in the parameter $\alpha$. The shape of the performance curve shifts rapidly as $\alpha$ varies, demonstrating that the estimation accuracy becomes highly dependent on the configuration of the beamforming strategy. In the extremely low SNR regime (for -30 dB and -20 dB sensing SNR), the CRB and ZZB diverge significantly. The ZZB, incorporating prior knowledge about the bound, deviates from the CRB, which does not account for prior information.  It also demonstrates that, in this region, variations in $\alpha$ does not has much effect on sensing, as the system’s performance is predominantly influenced by the prior information available rather than the current beamforming configuration.

\begin{figure}  
    \centering
    \includegraphics[width=0.95\linewidth]{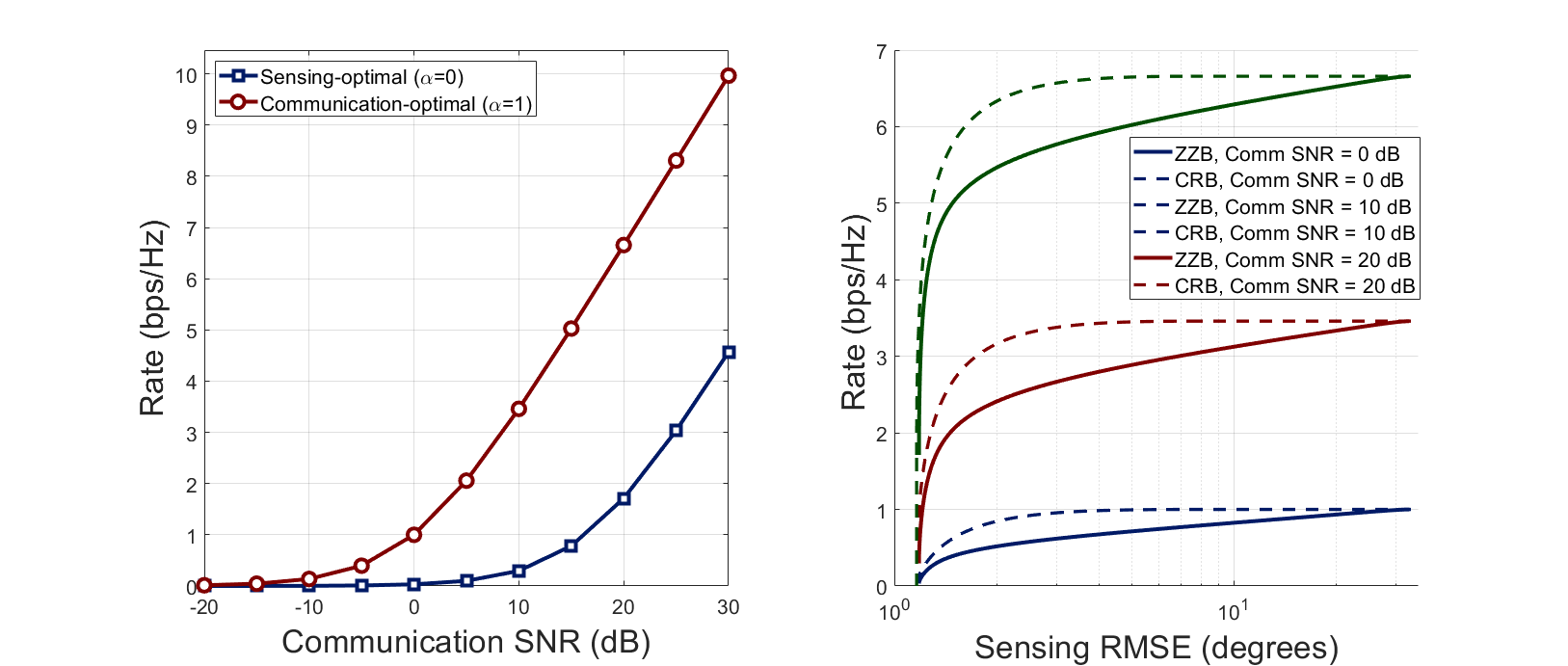}
    \caption{ Bistatic ISAC performance analysis and trade-offs. \textit{Left:} Communication rate versus SNR for sensing-optimal and communication-optimal beamforming. \textit{Right:} Rate-sensing pareto frontiers showing the fundamental trade-off between communication rate and sensing accuracy (ZZB error) at different communication SNR levels for fixed -10 dB Sensing SNR.}
    \label{fig:fig5}
\end{figure}

 Fig. \ref{fig:fig5} illustrates the pareto frontier for fixed Sensing SNR at threshold region. The left plot shows communication rate versus communication SNR for sensing optimal ($\alpha=0$) and communication-optimal ($\alpha=1$) beamforming strategies, demonstrating how performance diverges as communication SNR increases. The right plot presents the Pareto frontiers between communication rate and ZZB-based estimation RMSE at fixed sensing SNR of -10 dB while varying communication SNR across $\{0,10,20\}$ dB. At high communication SNR (20 dB), adjusting $\alpha$ offers significant control over the communication-sensing balance, creating a wide performance region. Conversely, at low communication SNR (-10 dB), the nearly flat frontier indicates minimal impact from beamforming adjustments as noise dominates performance. This confirms ZZB's effectiveness in capturing the diminishing returns of beamforming optimization at low SNR—a relationship that CRB fails to properly characterize. This also shows, traditional CRB at threshold region significantly overestimates system capabilities by predicting better performance than is practically achievable. ZZB provides a more realistic bound that avoids overestimating system capabilities when allocating resources.

\section{Conclusion} \label{section5}
In this paper, we presented a comprehensive analysis of bistatic ISAC systems where the sensing receiver has access only to the statistical properties of the received signal. Our focus was on characterizing the fundamental trade-off between communication rate and sensing accuracy, particularly in the a priori and threshold region. Through numerical evaluations, we showed that CRB-based performance assessments can be misleading in low SNR conditions. In contrast, the ZZB offers a more accurate performance bound by incorporating a priori information about the target's Angle of Arrival (AoA), especially in the threshold and asymptotic regions where noise, rather than beamforming configuration, dominates sensing performance. These findings suggest that system designers should prioritize the use of ZZB over CRB when evaluating performance trade-offs, particularly in passive sensing scenarios. Future work could extend this analysis to multiple access scenarios and investigate the impact of non-uniform prior distributions on system performance.

\bibliographystyle{IEEEtran}
\bibliography{bibliography}

\end{document}